\documentclass[twocolumn,showpacs,preprintnumbers,amsmath,amssymb]{revtex4}
\usepackage{graphicx}
\newcommand{\ltsim}{\protect\raisebox{-0.5ex}{$\:\stackrel{\textstyle <}{\sim}\:$}}
\newcommand{\gtsim}{\protect\raisebox{-0.5ex}{$\:\stackrel{\textstyle >}{\sim}\:$}}
\begin{document}
\title{Excitonic effect on optical response in one-dimensional two-band Hubbard model}
\author{H. Matsueda}
\email{matsueda@imr.tohoku.ac.jp}
\author{T. Tohyama}
\author{S. Maekawa}
\affiliation{Institute for Materials Research, Tohoku University, Sendai 980-8577, Japan}
\date{\today}
\begin{abstract}
Motivated by the gigantic nonlinear optical response in the halogen-bridged Ni-compounds, the underlying electronic states of the compounds are examined in the one-dimensional two-band Hubbard model, by studying the current-current correlation function and the charge density in the ground state. The dynamical density matrix renormalization group method is employed. We find that the low-energy peak of the correlation function consists of a single Lorentzian component for a parameter range appropriate to the compounds. This is due to an excitonic state induced by the intersite Coulomb repulsion between holes on the metal and halogen ions. This is consistent with the optical absorption spectra of the compounds. We suggest that the localization of holes on the metal ions in the ground state brings about the formation of the excitonic state.
\end{abstract}
\maketitle

One-dimensional Mott insulators such as cuprates and halogen-bridged Ni-compounds exhibit gigantic nonlinear optical responses, which have potential as opto-electronic devices.~\cite{Ogasawara,Kishida1,Kishida2,Mizuno1,Ono} It has been shown that this massive response arises because the photoexcited states across the Mott gap with even and odd parities are degenerate. This leads to the enhancement of the dipole matrix element of the third-order nonlinear susceptibility. In the Mott insulators, both the degeneracy and the optical gap reflect strong electron correlation.

The single-band Hubbard model with nearest-neighbor Coulomb repulsion at half-filling is a minimal model for the study of the optical response and photoexcited states in one-dimensional Mott insulators. The photoexcited states and optical response of the model have been examined by employing the dynamical density matrix renormalization group (DDMRG) method.~\cite{Matsueda,Jeckelmann1,Essler,Jeckelmann2,Jeckelmann3,Kancharla} The model and the method are suitable for the clarification of the general electronic properties of the insulators. However, the real compounds such as the cuprates and the halogen-bridged Ni-compounds are the insulators of charge transfer type, where both half-filled $d$-orbitals and empty $p$-orbitals for holes exist in one-dimensional networks of the transition metal and ligand ions. Therefore, the two-band Hubbard model is relevant for the insulators.

For the understanding of the optical response of the Ni-compounds with huge third-order nonlinear susceptibility, there are key electronic structure parameters in the two-band Hubbard model.~\cite{Ono,Fujimori,Okamoto1,Okamoto2} One of them is the charge transfer energy $\Delta$, which is related to the optical gap. In the Ni-compounds, $\Delta$ is estimated to be comparable with the $d$-$p$ hopping energy $t$ by the angle-resolved photoemission spectroscopy,~\cite{Fujimori} and the physical picture based on the perturbation expansion with respect to $t/\Delta$ may be violated. Another parameter is the intetsite Coulomb repulsion $V$ between holes on the metal and halogen ions, which causes an excitonic state. The presence of the excitonic states in the Ni-compounds has been confirmed through the optical absorption and photoconductivity measurements.~\cite{Ono,Okamoto1} Since the optical gap $\omega_{0}$ observed in the absorption spectrum is smaller than the threshold energy of the photoconductivity, the excitation by $\omega_{0}$ is not sufficient for photocarrier generation. This suggests that the excitonic state occurs. In fact, the optical absorption spectrum can be decomposed into a single Lorentzian component and a long tail structure.~\cite{Ono}

Motivated by the optical properties of the Ni-compounds, we numerically calculate the current-current correlation function in the two-band Hubbard model. A numerical simulation technique applied here is the DDMRG method.~\cite{Matsueda,White,Hallberg,Kuhner,Jeckelmann1,Essler,Jeckelmann2,Jeckelmann3,Kancharla} With changing $\Delta/t$ and $V/t$ systematically, we find that the low-energy peak of the correlation function consists of a single Lorentzian component for a parameter range appropriate to the compounds. This is due to the excitonic state induced by $V$, and consistent with the optical absorption spectra of the compounds. In addition to the correlation function, the charge density in the ground state also provides useful information to understand the formation of the excitonic state. Then, we calculate the charge density by using the DMRG method, and suggest that the localization of holes on metal ions in the ground state brings about the formation of the excitonic state.

The two-band Hubbard model on one-dimensional chain is given by
\begin{eqnarray}
H&=&\sum_{i}\epsilon_{i}n_{i}+\sum_{i,\sigma}t_{i,i+1}( c^{\dagger}_{i,\sigma}c_{i+1,\sigma} + {\rm H.c.} ) \nonumber \\
&& + \sum_{i}U_{i}n_{i,\uparrow}n_{i,\downarrow} + V\sum_{i}n_{i}n_{i+1} ,
\end{eqnarray}
where the operator $c^{\dagger}_{i,\sigma}$ creates a hole with spin $\sigma$ at site $i$, $n_{i}=n_{i,\uparrow}+n_{i,\downarrow}$ is the number operator, the level energy $\epsilon_{i}$ and the on-site Coulomb interaction $U_{i}$ are defined by $\epsilon_{i}=\Delta$ and $U_{i}=U_{p}$ for odd $i$ (halogen ion), and $\epsilon_{i}=0$ and $U_{i}=U_{d}$ for even $i$ (metal ion), respectively. $|t_{i,i+1}|=t$ and the sign of $t_{i,i+1}$ is taken from the electronic states of the Ni-compounds.~\cite{Kishida1} We introduce the intersite Coulomb interaction $V$ between holes on metal and neighboring halogen sites. For the charge transfer type insulators, $\Delta<U_{d}$. The number of holes is equal to that of metal sites. For a moment, the on-site Coulomb interaction at halogen sites, $U_{p}$, is neglected to make the roles of $V$ clear, although $U_{p}$ may be finite in the real compounds. In this paper, $U_{d}$ is fixed to be $U_{d}/t=8$.

We calculate the current-current correlation function defined by
\begin{equation}
\chi_{\gamma}(\omega)=-\frac{1}{\pi}{\rm Im}\left<0\right|j\frac{1}{\omega+E_{0}-H+i\gamma}j^{\dagger}\left|0\right> ,
\end{equation}    
where $j=i\sum_{i,\sigma}t_{i,i+1}( c^{\dagger}_{i,\sigma}c_{i+1,\sigma} - {\rm H.c.} )$ is the current operator, $\left|0\right>$ is the ground state with energy $E_{0}$, and $\gamma$ is a small number. We also calculate the charge density in the ground state, $N(i)=\left<0\right|n_{i}\left|0\right>$.

\begin{figure}
\begin{center}
\includegraphics[width=11cm]{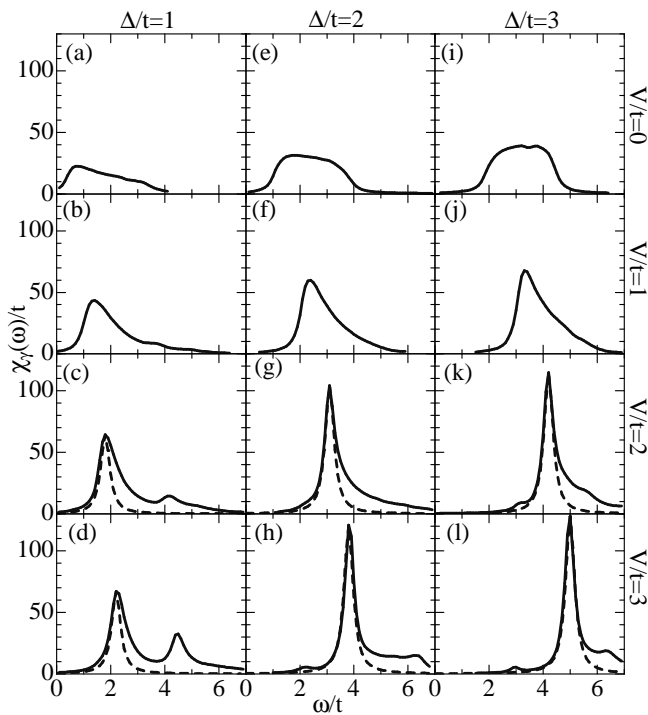}
\end{center}
\caption{$\chi_{\gamma}(\omega)$ as a function of $\Delta/t$ and $V/t$. The on-site Coulomb repulsion is fixed to be $U_{d}/t=8$ and $U_{p}/t=0$. Dashed lines denote single Lorentzian components.}
\end{figure}

Numerical simulations to obtain $\chi_{\gamma}(\omega)$ and $N(i)$ are separately carried out by the DMRG method under open boundary condition. The superblock is constructed by $129$ sites, i.e. $64$ unit cells plus one halogen site. It should be noted that both ends of a one-dimensional chain are halogen sites, since this configuration keeps the inversion symmetry of the system after the sweep procedure is converged. For any parameter set, the result for $129$ site system was compared with that for smaller clusters, so that a boundary effect inducing localization of electrons at chain edges can be identified. Additional impurity potentials on the two halogen ends are included by changing $H$ with $H-(V/4)(n_{1}+n_{L})$ to reduce the boundary effect. As will be mentioned below, the effect produces a small hump structure at around $\omega=\Delta$ for $V/t\ge 2$ in $\chi_{\gamma}$ even for the $129$ site system. For calculating $\chi_{\gamma}(\omega)$, the DDMRG method for the fixed broadening factor $\gamma/t=0.2$ is applied.~\cite{Matsueda,Kuhner,Jeckelmann2} To define the density matrix for mixed states in the DDMRG method, we take three target states, i.e. the ground state $\left|0\right>$, a photoexcited state $j^{\dagger}\left|0\right>$, and the correction vector $(\omega+E_{0}-H+i\gamma)^{-1}j^{\dagger}\left|0\right>$. For calculating $N(i)$, the standard DMRG method is applied.~\cite{White} For both $\chi_{\gamma}(\omega)$ and $N(i)$, the DMRG bases are truncated up to $m=300$.

In Fig. 1, $\chi_{\gamma}(\omega)$ is plotted for various values of $\Delta/t$ and $V/t$. A Lorentzian is also plotted in the Figure for $V/t\ge 2$. It is defined by $L_{\gamma}(\omega)=\gamma^{2}\chi_{\gamma}(\omega_{0})/[(\omega-\omega_{0})^{2}+\gamma^{2}]$ with $\omega_{0}$ denoting the excitation energy which provides the maximum $\chi_{\gamma}(\omega)$. We focus on whether the full width of half maximum for a low-energy peak of $\chi_{\gamma}(\omega)$ is equal to that for the Lorentzian. As seen in Fig. 1, the width of the spectrum for $\Delta/t=1$ is qualitatively different from that for $\Delta/t=2$. (The spectrum for $\Delta/t=3$, which corresponds to a parameter for the cuprates, is similar to that for $\Delta/t=2$.)

Let us start with $\chi_{\gamma}(\omega)$ for $\Delta/t=2$ [see Figs. 1(e), (f), (g), and (h)]. For $V/t=0$, the spectra are composed of two broad structures. One is a broad peak at around $\omega/t=1.5$. This is due to the excitation process from the Zhang-Rice singlet bound state to the upper Hubbard band~\cite{Mizuno2}, which we call the process (A). The other is a shoulder at around $\omega/t=3$. This is due to the process from the unbound states to the upper Hubbard band, which we call the process (B). The energy differences between the two structures can be scaled by the coupling constant of the singlet pairing, $J_{0}=4t^{2}[(1/\Delta)+(1/(U_{d}-\Delta))]$. With increasing $V$, the spectral intensity due to the process (A) increases, while the intensity due to the process (B) rapidly decreases. For $V/t\ge 2$, a main peak due to the process (A) is located at $\omega_{0}\sim\Delta+V-t$. The full width at half maximum for the main peak is almost the same as that for the single Lorentzian. Therefore, we find that the excitonic state occurs for $V/t\ge 2$. It is noted that a small hump structure seen around $\omega=2=\Delta$ in Fig. 1(h) is due to the boundary effect. In Fig. 2, $\chi_{\gamma}(\omega)$ is replotted for various $V$ with fixing $\Delta/t=2$. $\chi_{\gamma}(\omega_{0})$ is enhanced by increasing $V$ for $2\le V/t \le3$, while it tends to be suppressed with larger $V$.

\begin{figure}
\begin{center}
\includegraphics[width=10cm]{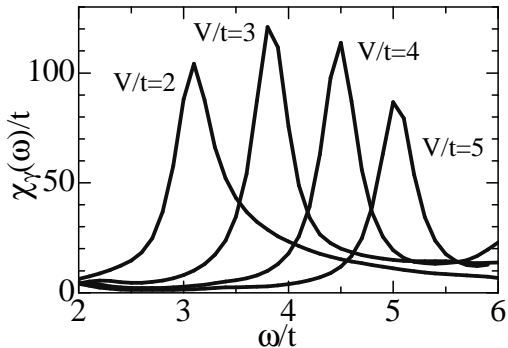}
\end{center}
\caption{$V$ dependence of $\chi_{\gamma}(\omega)$ for $\Delta/t=2$.}
\end{figure}

Next, we examine $\chi_{\gamma}(\omega)$ for $\Delta/t=1$ [see Figs. 1(a), (b), (c), and (d)]. For $V/t\le 1$, the intensity due to the process (A) increases with increasing $V/t$. The intensity due to the process (B) is almost kept. For $V/t\ge 2$, the main peak structure is observed near $\omega_{0}\sim\Delta+V-t$. However, it is not clear whether the excitonic state is formed or not, since the full width at half maximum for the main peak is larger than that for the single Lorentzian. Even though $V$ is further increased, the fitting does not works well. Furthermore, for $V/t>3$, $\chi_{\gamma}(\omega_{0})$ decreases with increasing $V$. The change of $\chi_{\gamma}(\omega)$ is larger than that for $\Delta/t=2$ (not shown). An additional peak structure due to the process (b) is also observed near $\omega/t=4$ in Fig. 1(c) and $\omega/t=4.5$ in Fig. 1(d). The intensity of the additional peak increases with increasing $V$.

\begin{figure}
\begin{center}
\includegraphics[width=10cm]{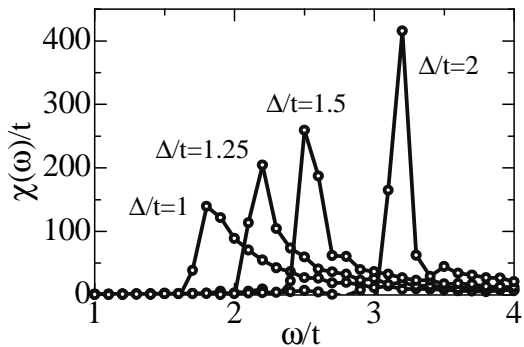}
\end{center}
\caption{$\Delta$ dependence of $\chi(\omega)$ for $V/t=2$.}
\end{figure}

As discussed in the previous two paragraphs, the excitonic state seen as a low-energy peak of $\chi_{\gamma}(\omega)$ depends on the magnitude of $\Delta/t$. Let us examine how the excitonic state is destroyed with decreasing $\Delta/t$. Since, in the DDMRG method, the spectral shape depends on the magnitude of the broadening factor $\gamma$, we remove the effect of the broadening by using the Lorentz transformation.~\cite{deconvolution} After the removal of the effect, $\chi_{\gamma}(\omega)$ is expressed by $\chi(\omega)$. In the actual numerical calculation, $\chi(\omega)$ still has a finite broadening factor $\delta\omega ( < \gamma )$. The factor is set to be $\delta\omega=\gamma/2$.~\cite{deconvolution} Figure 3 shows the $\Delta$ dependence of $\chi(\omega)$ for $V/t=2$. The peak intensity linearly decreases with decreasing $\Delta/t$. For $\Delta/t\ge 1.5$, the full width at half maximum of the main peak is nearly twice the broadening factor, $2\delta\omega$. The width becomes narrower if we choose $\gamma$ smaller than $0.2t$. This indicates the presence of the excitonic state. However, the width for $\Delta/t\le 1.25$ is beyond $2\delta\omega$. The width increases with decreasing $\Delta/t$ for $\Delta/t\le 1.25$. Even for $V/t\ge3$, the width for $\Delta/t\ltsim 1.25$ is still beyond $2\delta\omega$ (not shown).

\begin{figure}
\begin{center}
\includegraphics[width=8cm]{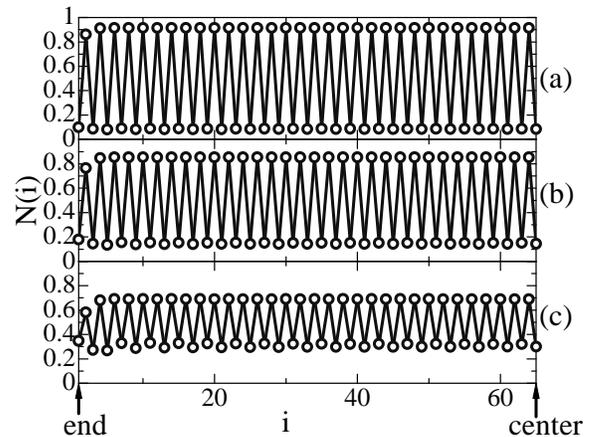}
\end{center}
\caption{Charge density from one halogen end ( $i=0$ ) to the central halogen site ( $i=65$ ) for $V/t=2$. (a) $\Delta/t=3$, (b) $\Delta/t=2$, and (c) $\Delta/t=1$.}
\end{figure}

In the following, we show that the broad spectrum for $\Delta/t\ltsim 1.25$ is caused by delocalization of holes in the ground state. Figure 4 shows the charge density in the ground state for $\Delta/t=3$, $2$ and $1$ with $V/t=2$.~\cite{caption} The differences of the charge density between metal and halogen ions for $\Delta/t=3$, $2$ and $1$ are $0.83$, $0.71$ and $0.38$, respectively. In contrast with the difference for $\Delta/t=3$ and $2$, this is small for $\Delta/t=1$. The delocalization of holes for $\Delta/t=1$ is related to the destruction of the excitonic state: As seen in Fig. 3, the low-energy peak of $\chi(\omega)$ for $V/t=2$ consists of a single Lorentzian component for $\Delta/t\gtsim 1.5$, while does not for $\Delta/t\ltsim 1.25$. Therefore, the larger the charge localization is, the more stable the excitonic state is. In other words, holes should be localized on metal ions in the ground state, when the excitonic state is formed.

Finally, $\chi(\omega)$ is compared with the optical absorption spectrum of the Ni-compounds. For $\Delta/t\ge 1.5$ and $V/t=2$ in Fig. 3, $\chi(\omega)$ is consistent with the optical absorption spectrum observed experimentally: $\chi(\omega)$ has a single Lorentzian peak followed by a long tail structure.~\cite{Ono} For these parameter values, the excitation energy $\omega_{0}$ which corresponds to the optical gap is estimated to be $\omega_{0}\gtsim 2.5t\sim 2.5({\rm eV})$, if we take $t\sim 1({\rm eV})$.~\cite{Fujimori} Compared with the observed optical gap [$1.3({\rm eV})\sim 2.0({\rm eV})$], $\omega_{0}$ may be overestimated. However, we have confirmed numerically that the finite $U_{p}$ extends the region where the excitonic state occurs. For the realistic $U_{p}$ and $\Delta/t\gtsim 1$, $\omega_{0}$ becomes consistent with the observed optical gap.

In summary, we have investigated the current-current correlation function and the charge density of one-dimensional two-band Hubbard model by using the density matrix renormalization group approach. In the two-band Hubbard model, the intersite Coulomb interaction $V$ between holes on metal and halogen ions is a primary factor to produce the exciton effect on the correlation function. We found that the low-energy peak of the correlation function consists of a single Lorentzian component for $\Delta/t\gtsim 1.5$ and $V/t\ge 2$ in the case that $U_{d}/t=8$ and $U_{p}=0$. This means the formation of the excitonic state. This is consistent with the optical absorption spectrum observed in the Ni-compounds. By examining the correlation function and the charge density in the ground state, we suggested that the localization of holes on metal ions in the ground state brings about the formation of the excitonic state.

This work was supported by NAREGI Nanoscience roject and Grant-in-Aid for Scientific Research from the Ministry of Education, Culture, Sports, Science and Technology of Japan, and CREST. A part of the numerical calculations was performed in the supercomputing facilities in ISSP, University of Tokyo and IMR, Tohoku University.


\begin{thebibliography}{99}
\bibitem{caption}
There exists small charge modulation on the halogen sites. However, the modulation is caused by the presence of the free boundary.
\end{thebibliography}
\end{document}